\documentclass[12pt]{article}
\usepackage{amsfonts,epsfig}
\newcommand{\beq}{\begin{eqnarray}}
\newcommand{\eeq}{\end{eqnarray}}
\begin{document}
\title{Supernovae, Landau Levels, and Pulsar Kicks}
\author{Leonard S. Kisslinger, Department of Physics, Carnegie Mellon University}
\date{}
\maketitle
\begin{abstract} We derive the energy asymmetry given the proto-neutron
star during the time when the neutrino sphere is near the surface of the
proto-neutron star, using the modified URCA process. The electrons produced
with the anti-neutrinos are in Landau levels due to the strong magnetic
field, and this leads to asymmetry in the neutrino momentum, and a pulsar
kick. Our main prediction is that the large pulsar kicks start at about 10 s
and last for about 10 s, with the corresponding neutrinos correlated in the
direction of the magnetic field.
\end{abstract}

\section{Outline}

\hspace{5mm} {\bf Supernova gravitational collapse of massive star often  
creates a pulsar.}
\vspace{1mm}

{\bf Many pulsars are moving with velocities much larger than stars in
our galaxy: PULSAR KICK}
\vspace{1mm}

{\bf I'll review hydrodynamics of the collapse and  neutrino emission
$0<t<10$ s:  NO GO}
\vspace{1mm}

{\bf I'll review sterile neutrinos possibly seen at LSND (\& MiniBooNe ?) 
giving the pulsar kicks.}
\vspace{1mm}

{\bf Research by Ernest M. Henley, Mikkel B. Johnson and Leonard S. 
Kisslinger: We get pulsar kicks consistent with observations
from asymmmetry due to the electrons produced in the modified URCA process
in strong magnetic fields being in Landau levels}

\section{Review Supernovae and Neutron Star 
Formation}

  A supernova is the gravitational collapse of a massive star ($\geq~$ 8 sun
  masses):
\vspace{1mm}

  1. Collapse to density $> 10^{14}$ g cm$^{-3} >$ nuclear density

\hspace{6mm} Shocks, bounce, etc.
\vspace{1mm}

2. Protoneutron star formed $\sim$ 0.01 s. 

\hspace{6mm}Neutrinos trapped in neutrinosphere. Radius of neutrinosphere 
$\sim$ 40 km.

\clearpage

\begin{figure}[h!]
\begin{center}
\epsfig{file=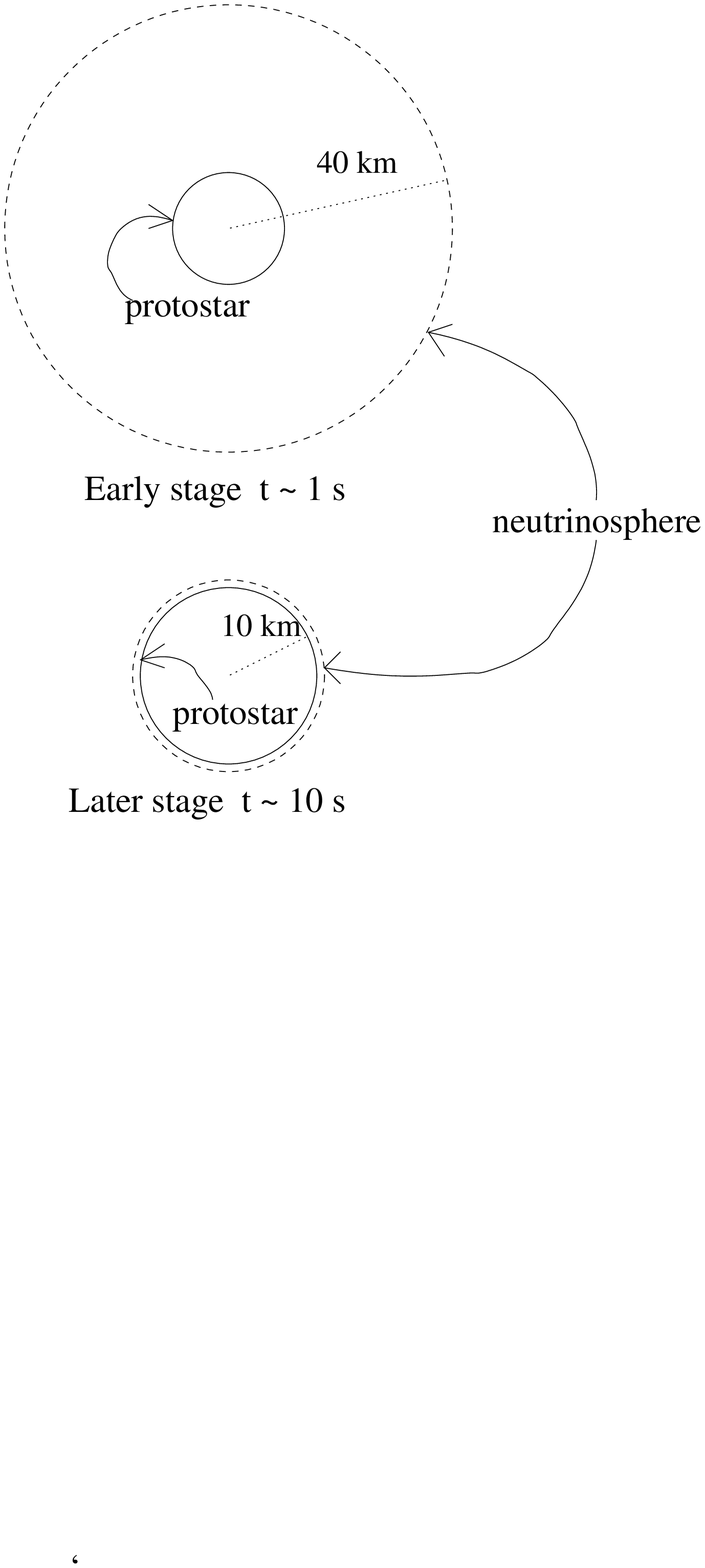,height=18cm,width=9cm}
%\caption{}
%{\label{Fig.4}}
\end{center}
\end{figure}

\vspace{-8cm}

3. From 0.1 to 10 sec neutrinosphere contracts from $\sim$ 40 km to protostar
radius $\sim$ 10 km

\hspace{8mm}Neutrinos carry gravitational energy from the emerging star.

4. From $\sim$ 10 to $\sim$ 50 sec n-n collisions dominate neutrino
production and protoneutron star cooling. 

\hspace{8mm}The modified URCA process dominates energy emmision by neutrinos.
\beq
           n + n &\rightarrow& n + p + e^- +\bar{\nu}_e \nonumber
\eeq

\subsection{Pulsar Kicks}

{\bf Resulting neutron stars (pulsar) have a bimodal velocity distribution}
\beq
            0 &<& v\;<\; 300 {\rm km\;s^{-1}} \nonumber \\
          1000 &<& v\;<\; 1500 {\rm km\;s^{-1}}\; \nonumber
\eeq
(see, e.g., B.M.S. Hansen and E.S. Phinney, astro-ph/9708071) 

  I.e., pulsars with high luminoscity tend to have velocities more than
an order of magnitude greater than star velocities in our galaxy.

\newpage

\subsection{Attempts at Explanation}

   There have been many papers written on attempts to explain the pulsar
kicks by processes taking place during the first 10 s, when most of the
energy from the collapse is carried off by neutrino emission. The main
attempts have been asymmetries in the hydrodynamic processes associated
with the collapse, the scattering of neutrinos produced by the URCA process
by nucleons polarized by the strong magnetic field within the neutrinosphere,
and by sterile neutrino processes. We briefly review these attempts.

\subsubsection{ Asymmetries in core collapse} There have been many attempts
to explain the pulsar kicks by the asymmetries in the initial collapse
of the massive star. Recently authors of a calculation postulating a slow
collapse (t $\geq$ 1s) claim that they can predict pulsar velocities 
greater than 500 km/s  (L. Scheck et. al., P.R.L. 92 (2004) 011103).
In a review by J.Murphy, A. Burrows, and A Heger, Astrophysics J.615 
(2004) 615, a careful analysis of all hydrodynamic calculations shows
that one cannot obtain a kick of more than about 200 km/s.

\subsubsection{ Processes involving standard neutrinos with a strong 
magnetic field}  During the first 10 s, when the neutrinosphere has
a radius starting about 40 km, the URCA process dominates neutrino
emission (see O.F. Dorofeev et al, Sov. Astron. Lett. 11 (1985)),
\beq 
   n&\rightarrow& p +e^- \bar{\nu} \nonumber \\
   e^- + p &\rightarrow& n + \nu_e \nonumber \\
   e^+ + n &\rightarrow& p + \bar{\nu}_e \; , \nonumber
\eeq
where the nucleons are polarized by the strong magnetic fields of the
protoneutron star. A few years ago calculations of elastic neutrino scattering
from the polarized nucleons ( C.J. Horowitz and J. Piekarewitz, N.P. A640 
(1998) 281; C.J. Horowitz and G. Li, PRL 80 (1998) 3694) found this process
to be promising to explain the pulsar kicks.

  However, if one includes the neutrino absorbtion opacities modified by the
strong magnetic fields, the asymmetric neutrino emission is reduced, and
these processes with standard neutrinos cannot account for the pulsar kicks,
D. Lai and Y-Z. Qian, astro-ph/9802345, ApJ.(1998). Note that although the
neutrinosphere at 1 s is about 40 km, the mean free path of neutrinos is
only or the order of 1 cm, and the asymmetric neutrinos are not emitted.
\vspace{5mm}

{\bf Mechanisms with standard neutrinos during the first 10 s have not been 
able to account for the large pulsar velocities.} 

\clearpage

\subsubsection{Pulsar kicks from sterile neutrinos in a strong magnetic field}

If sterile neutrinos exist they can be produced by oscillations in
the protoneutron star matter. In the presence of the strong magnetic fields
of the protoneutron star neutrino asymmetries will be produced. Since the
opacities of sterile neutrinos are very small, this has been studied as
a possible source of the pulsar kicks (A Kusenko and G. Segre, PRL 77 (1996) 
4872; PL B396 (1997) 197.

Using an extension of  usual concept of the CKM mixing matrix giving the
weak interaction states in terms of the mass eigenstates, sterile/active 
neutrino mixing given by mixing angle $|\theta_m| << 1$:
\beq
      |\nu_1> &=& cos\theta_m |\nu_e> -sin\theta_m |\nu_s> {\rm trapped}
\nonumber \\
      |\nu_2> &=& sin\theta_m |\nu_e> +cos\theta_m |\nu_s> {\rm not\; trapped}
 \nonumber
\eeq 

  In a calculation that requires the consistency of sterile neutrinos and
dark matter (G.A Fuller, A. Kusenko, I. Mocioiu and S. Pascoli, PR D68 (2003) 
103002) the parameters needed to give the kick were found, as illustrated
in the figure below
\begin{figure}[h]
\begin{center}
\epsfig{file=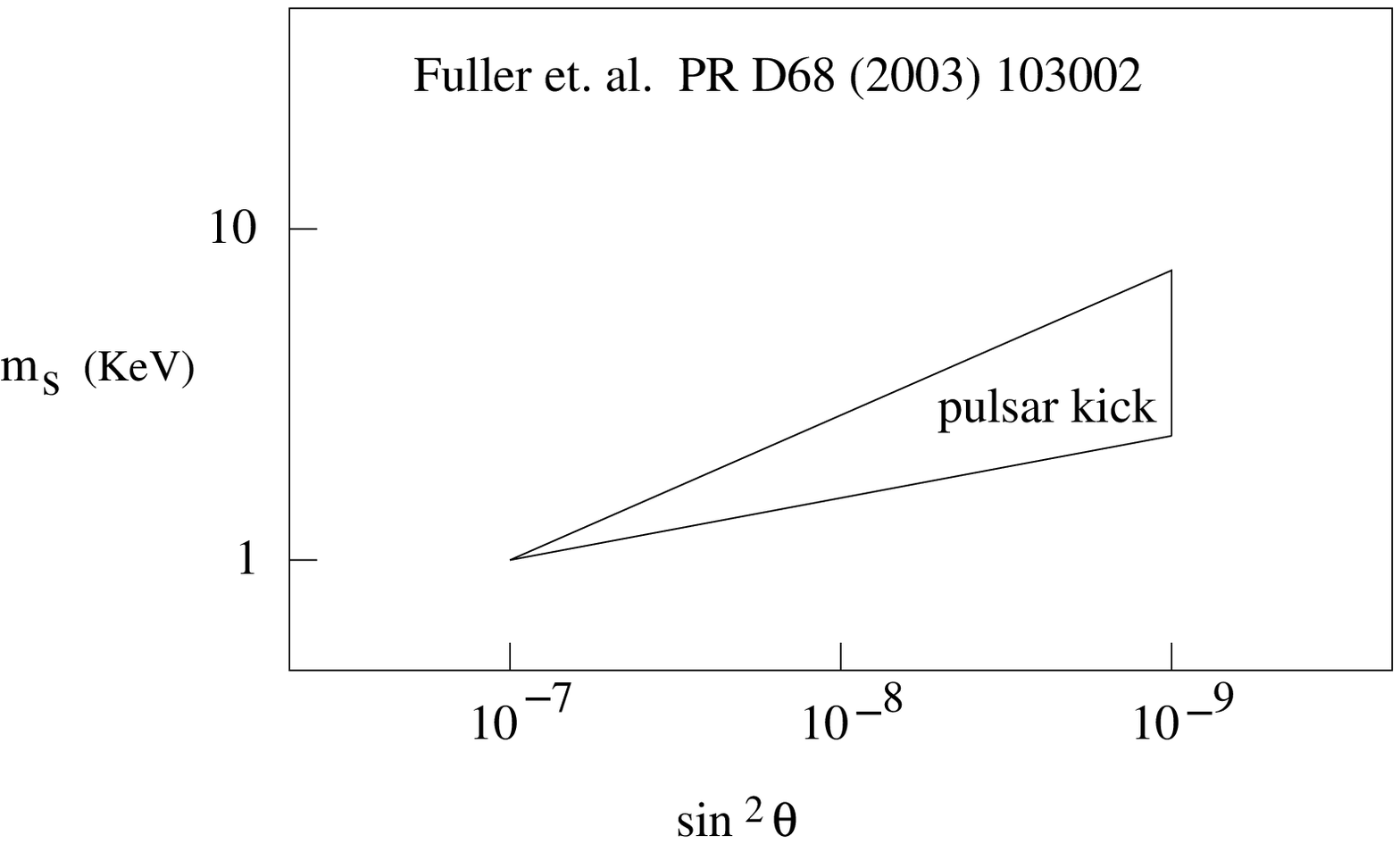,height=3in,width=6in}
%\label
\end{center}
\end{figure}

One sees that a sterile neutrino mass of 1- 10 Kev is needed, which is
not expected. 
\vspace{5mm}

{\bf New data on the sterile neutrino's mass and mixing angle is expected
from MiniBoone. If the results are similar to those of LSND, with fits
such as $\Delta m^2 = 0.2 ev^2\;and\;sin^2\theta=0.068$ (From William 
Lewis, LANL) there is a possibility to get the pulsar kicks from sterile
neutrinos.

\clearpage

\section{Pulsar Kicks From the Modified URCA Process
in a Strong Magnetic Field}

 In the later stage ($t > \sim 10 s$) the Modified URCA process:
\beq
\label{1}
    n + n &\rightarrow& n + p + e^- + \bar{\nu}_e 
\eeq
dominates the cooling of the neutron star.
\begin{figure}[ht]
%\begin{center}
\epsfig{file=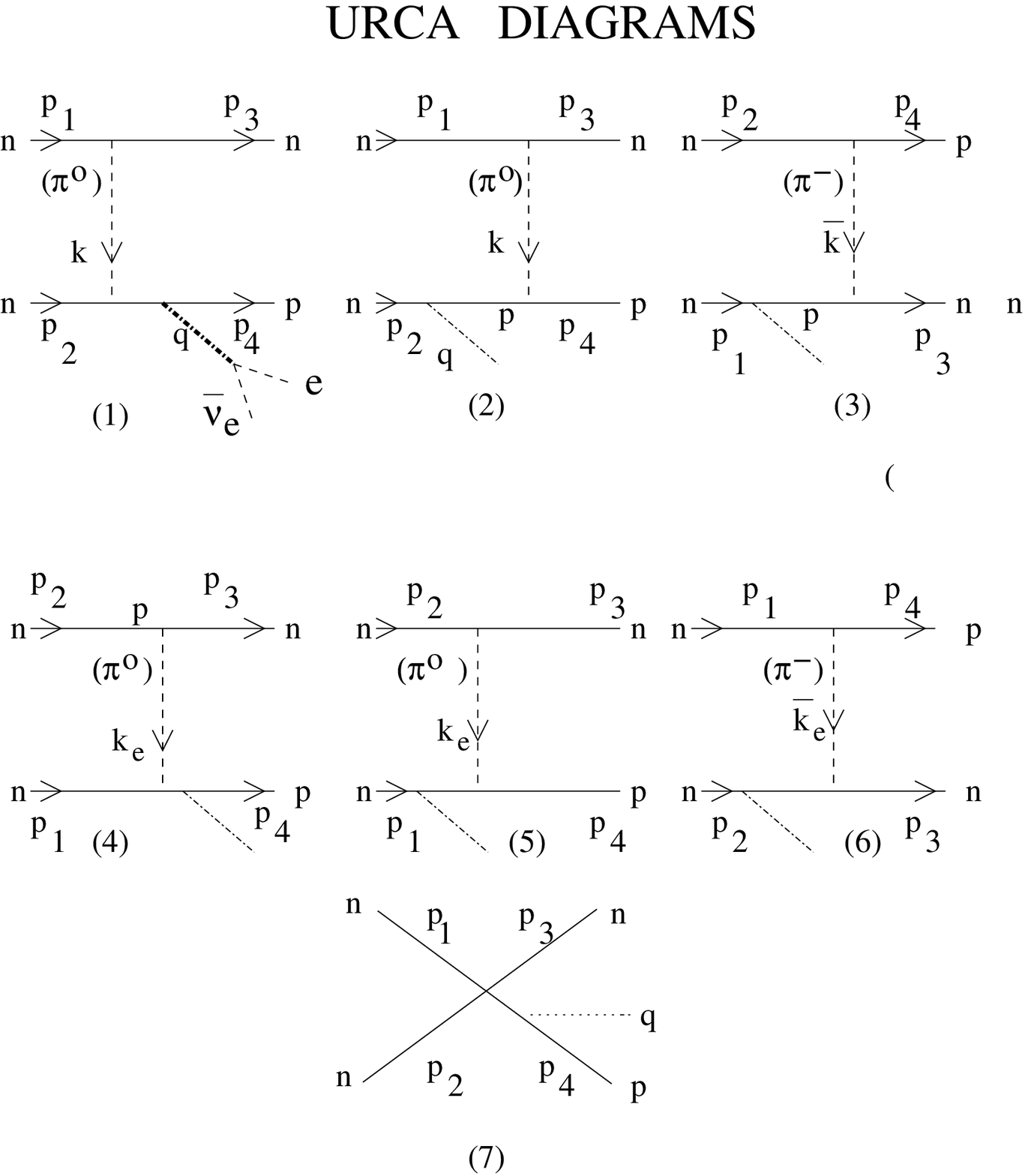,height=12cm,width=15cm}
%\caption{}
%{\label{Fig.4}}
%\end{center}
\end{figure}

 The early work includes

J.N. Bahcall and R.A. Wolf, PRL 14 (1965) 343; PR 140 (1965) B1452;

E.G. Flowers, P.G. Sutherland and J.R. Bond, PRD 12 (1975) 315.

B.L. Friman and O.V. Maxwell, Astrophysical J. 232 (1979) 541, who used
pion exchange and short-range potential in detailed calculations.
\newpage

Modified URCA ingredients: Nonrelativistic OPE interaction and Standard 
W-S Model:}
\beq
\label{2}
   V_{OPE} &=& (\frac{f}{m_\pi})^2 \sigma^{(1)}\cdot k\frac{-1}{k^2 +m_{\pi}^2}
 \sigma^{(2)}\cdot k \tau^{(1)}\cdot \tau^{(2)} \; ,  
\eeq
with $f \simeq 1.0$, $\sigma,\tau$ the Pauli spin, isospin operators, and
(1,2) refer to the two nucleons at the pion vertices.

 Weak axial n-p vertex in nonrelativistic Standard Weinberg-Salam model:
\beq
\label{3}
 {\rm Weak\; Axial\;\;}W_A &=& -\frac{G}{\sqrt{2}} g_A  \chi_p^\dagger 
\vec{l}\cdot \sigma \chi_n \; ,
\eeq 
{\bf with G=$\frac{10^{-5}}{M_p^2},g_A = 1.26 $, and the lepton current 
defined by}
\beq
       l_\mu &=& \bar{u}(q^e)\gamma_\mu(1-\gamma_5)u(q^\nu) \nonumber \; .
\eeq
 We need the matrix element product
\beq
   \Sigma_{lepton,nucleon\;spins}|M|^2 &\equiv& \Sigma_{lepton,nucleon\;spins}
|V_{strong}W_{weak}|^2 \nonumber \;.
\eeq
 
In previous research the vector-axial vector product was used. However,
we realize that in a strong magnetic field the axial-axial product, which
is much larger, can be used, as the magnetic field provides the asymmetry.

We also make use of the fact that in the strong B field with the 
electrons in Landau levels, $u(q_e)\rightarrow \psi^{Landau}(q_e)$
one can get a large pulsar kick without n-polarization, greatly simplifying 
the calculation. The stationary wave function of an electron in a strong 
magnetic field is (J.J. Matese and R.F. O'Connel,  Phys. Rev. 180, 1289 (1969))
\beq
\label{4}
         \vec{B} &=& B \hat{z} \nonumber \\
     \psi^{Landau,n=o}_{\vec{p}}(q^e_\perp,p_z,\phi) &=& i(\sqrt{\gamma})^{-1}
e^{-(q^e_\perp)^2/2 \gamma} \nonumber \\
    &&\psi^{Dirac,-}(q^e) \delta(q^e-p^e_z) \; ,
\eeq
where $\gamma = B m_e^2/2B_c$, with the critical magnetic field $B_c\simeq
4\times 10^{13} G$. I.e., in a strong magnetic field the electron moves as 
a plane wave in the z direction, with energy of Landau levels, and is fully 
negatively polarized in the lowest Landau level.

The leptonic traces needed are
\beq
\label{5}
   Tr(lepton)[l^{\dagger}_i l_j]&=& Tr[l^{\dagger}_i l_j](\alpha)
+Tr[l^{\dagger}_i l_j](\beta)  \nonumber \\ 
 &&+Tr[l^{\dagger}_i l_j](\gamma) \nonumber \\ 
 Tr[l^{\dagger}4\gamma_i l_j](\alpha) &=& -4\gamma^{-1} g_{ij}q^e\cdot q^\nu
(1 + (1-\delta_{i3}))
 \nonumber \\
  Tr[l^{\dagger}_i l_j](\beta) &=& i4\gamma^{-1} \epsilon_{ijk}q^\nu_k q^e_0
\nonumber \\
  Tr[l^{\dagger}_i l_j](\gamma) &=& 4\gamma^{-1} (q^e_i q^\nu_j+q^e_j q^\nu_i)
 \; .
\eeq

Our notation for the matrix element products is
\beq
\label{6}
         k&=& p_1 -p_3\;,\;p=p_2-p_1\;,\;;k^e=k+p \nonumber \\
        A &=& (\frac{f}{m_\pi})^2\frac{G}{\sqrt{2}}g_A \frac{1}{\omega}\frac{1}
{(k^2+m_\pi^2)} \nonumber \\
    R(k)&=& \frac{k^2+m_\pi^2}{(k^{e})^2+m_\pi^2}\; .
\eeq

\subsection{Neutrino emissivity}

Defining the  Axial Product Matrix Element, $M_{AA} \equiv |M_A|^2$:
\beq
\label{7}
 M_{AA} &=&  | Tr(lepton)[l^{\dagger}_i l_j [W_A(V_{OPE} \nonumber \\
  &&+ exchange)]_{ij}|^2\; ,
\eeq 
the  general form of neutrino emissivity with A-A proces is
\beq
\label{8}
   e^\nu &=& (2\pi)^4 \int \prod_{i=1}^{4}\frac{d^3 p^i}{(2 \pi)^3}
\frac{d^3 q^e}{2\omega^e (2 \pi)^3}\frac{d^3 q^\nu}{(2 \pi)^3} 
\delta(E_{final}\nonumber \\
  &&-E_{initial}) \delta(\vec{p}_{final}-\vec{p}_{initial}) M_{AA} \mathcal{F}
 \; ,
\eeq
where $\mathcal{F}$ is the product of the initial and final Fermi-Dirac 
functions corresponding to the temperature and density of the medium.
$\int \frac{d^3 q^\nu}{(2 \pi)^3}$ omitted in present work. 

The nucleons and the electrons are in
thermal equilibrium. A crucial observation of this work is that the time 
scale of the strong interaction in the modified URCA process 
is short compared to the time scale for the electron, which is that of 
electromagnetic interaction. Therefore the proton 
quickly reaches thermal equilibrium, while the process of the electron
state transforming to the n=o Landau state does not intefere with the
proton reaching its Fermi momentum. Therefore we can use the values
for the nucleon and electron momenta in the matter of the protoneutron
star, as derived in the Friman-Maxwell paper.

 The $\int d^3 q^e$, involving a Landau level, makes use of
the momentum space representation, Eq. (\ref{4}), and the
integral over the transverse momentum direction is thereby given as
\beq
\label{9}
       \int dq^e_\perp q^e_\perp e^{- (q^e_\perp)^2/2\lambda} &=& 2\lambda
 \; .
\eeq
From this and Eq.(\ref{5}) one finds that the factors of $\gamma$
from the Landau level electrons cancel. 

  Recognizing that the nucleons are in thermal equilibrium, with Fermi
momentum $p_F$, the
angular integrals can be done using the the pion exchange momentum in the
direct diagram, as an independent vector by inserting: 
\beq
\label{10}
        \int d^3 \delta(\vec{k}-\vec{p}_1+\vec{p}_3)&=& I \; .
\eeq
For the D-D diagram the only angular integral needed is $d\Omega_k$, but
for the E-E and D-E terms two angular integrals are needed. We use
$d\Omega_{k^e}$, which is almost independent, and also $d\Omega_p$ as a
check. The $d\Omega_{q^e}$ is modified by a factor from the standard result.
\clearpage

  With no nucleon polarization the calculation is greatly simplified, and
we find 
\beq
\label{11}
       \int\int M_{A-A} &=& 1.38 \times 10^4 A^2 q^e q^\nu_z p_F^4 
\;{\rm \;(k,k^e)\;formalism}\nonumber \\
      \int\int M_{A-A} &=& 3.8 \times 10^4 A^2 q^e q^\nu_z p_F^4 
\;{\rm \;(k,p)\;formalism}
\eeq

The result for the energy integral over the Fermi distributions is that of
Friman-Maxwell:
\beq
\label{12}
    I&=& 5.68 \times 10^3 (kT)^8
\eeq
Using the standard values for the nucleon Fermi momenta, and  $q^e=85 MeV$, 
 $q^\nu = 4.7 kT$ we find the emissivity
\beq
\label{13}
  \epsilon^{AS}(k,k^e) &\simeq& 1.18\times 10^{24} (\frac{T}{10^9 K})^7 
{\rm \;erg\;cm^{-3}s^{-1}} \nonumber \\
  \epsilon^{AS}(k,p) &\simeq& 2.96\times 10^{24} (\frac{T}{10^9 K})^7 
{\rm erg \;cm^{-3}s^{-1}} \;,
\eeq
in the two formalisms.

 During $10s< t < 30s$, $10^9K<T<10^{11}K$ approximate period and
temperature, the mean free path of the neutrinos emitted is or the
order of a few cm. Thus the modified URCA asymmetric emission takes
place when the neutrinosphere is just within the neutron star, as depicted.
\begin{figure}[h]
\begin{center}
\epsfig{file=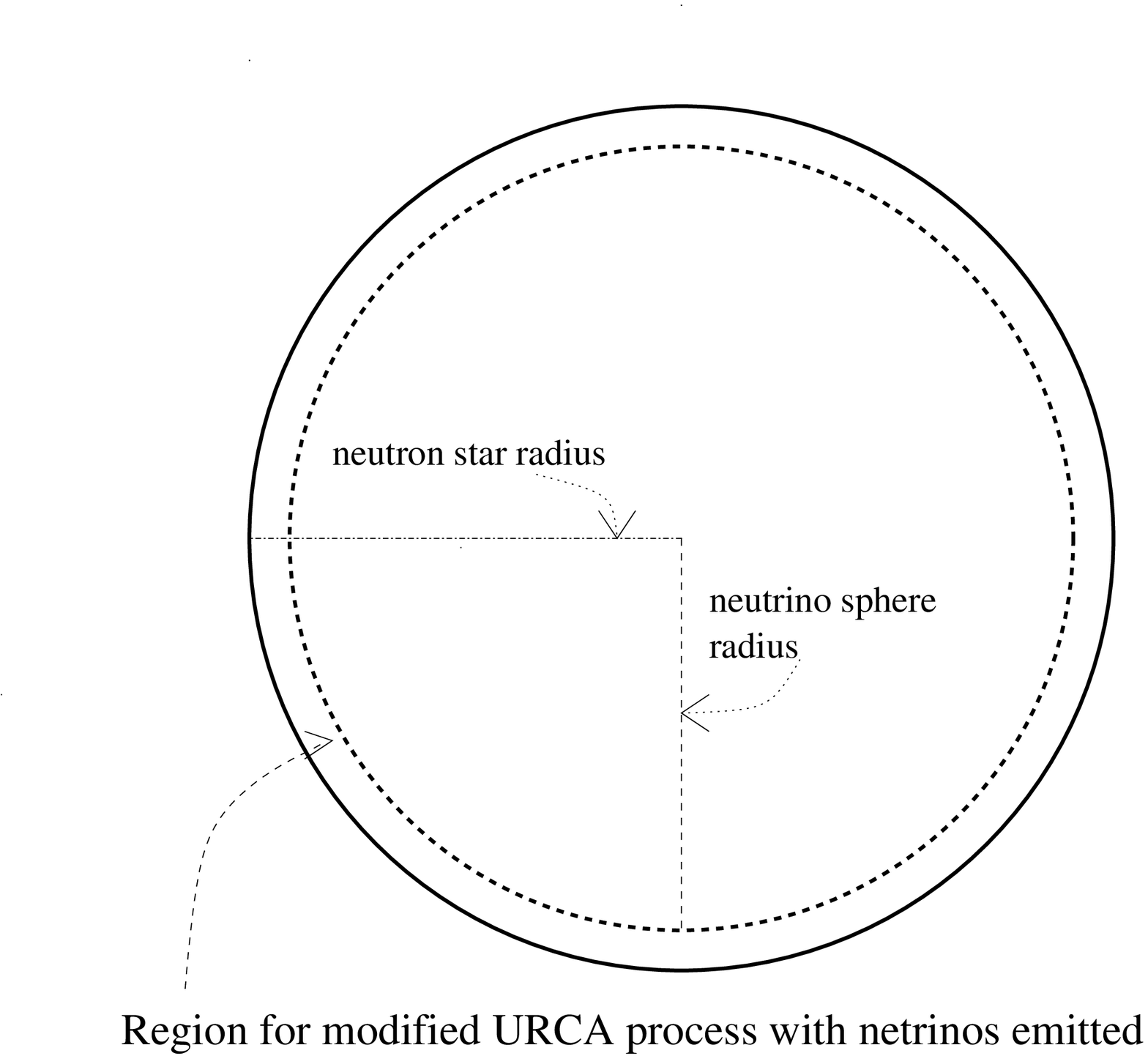,height=5cm,width=6cm}
%\caption{}
%{\label{Fig.2}}
\end{center}
\end{figure}
Therefore, the asymmetric energy emitted in the 10-20 s period is
\beq
\label{14}
   E^{AS}(k,k^e)&\simeq& 1.21 \times 10^{40}(\frac{T}{10^9 K})^7 
(R_{ns}^3-R_\nu^3) {\rm \;ergs} \nonumber \\
   E^{AS}(k,p)&\simeq& 2.74 \times E^{AS}(k,k^e) \; , 
\eeq
with a factor of .52 for the neutrinos that miss the protoneutron star.
\clearpage
\subsubsection{Radius of neutrinosphere during modified URCA emission}

   The final step in our derivation is to estimate the volume in which the
neutrino emission takes place with the modified URCA process in a strong
magntic field, which means finding the radius of the neutrinosphere during
the period of emission. 

  Our starting point is the energy-momentum tensor, $T^{\mu \nu}$ for the 
neutrinos with a distribution function $f_\nu(\vec{x},\vec{k},t)$ for each
type of neutrino, giving an energy density,$U$ and momentum density $\vec{F}$
\beq
\label{15}
    U&=& T^{00}=\int \frac{d^3k}{(2\pi)^3} k_0 f_\nu \nonumber \\
    F^i &=& T^{0i}= \int \frac{d^3k}{(2\pi)^3} k^i f_\nu \; .
\eeq
Using the Conservation of momentum, $\partial_\nu T^{0 \nu} = 0$ gives 
t-dependence:
\beq
   \partial_t U + \Delta \cdot \vec{F}& =& 0 \nonumber \; .
\eeq
Based on these equations, and using the Spherical Eddington Model for neutrino
atmosphere (Schindler-Shapiro,ApJ 259 (1982) 311; Janka-Raffelt, PR D59 (1998) 
023005, Barkovich-Olivo-Montemayor, hep-ph/0503113) with techniques that can
be found in Barkovich et al, we find for the radius of the neutrinosphere
during the period of emission:
\beq
\label{16}
      R_\nu &\simeq& 9.96 {\rm \;km} 
\eeq
when when the temperature is in the range $T\simeq 10^{10} K$ giving
\beq
\label{17}
     E^{AS} &\simeq& 1.45 \times 10^{42} {\rm ergs} (\frac{T}{10^9})^7 \; ,
\eeq
and a pulsar velocity v = 1000 km/s for an expected T=10$^{10}$ K. Our
results are as shown:
\begin{figure}[h]
\begin{center}
\epsfig{file=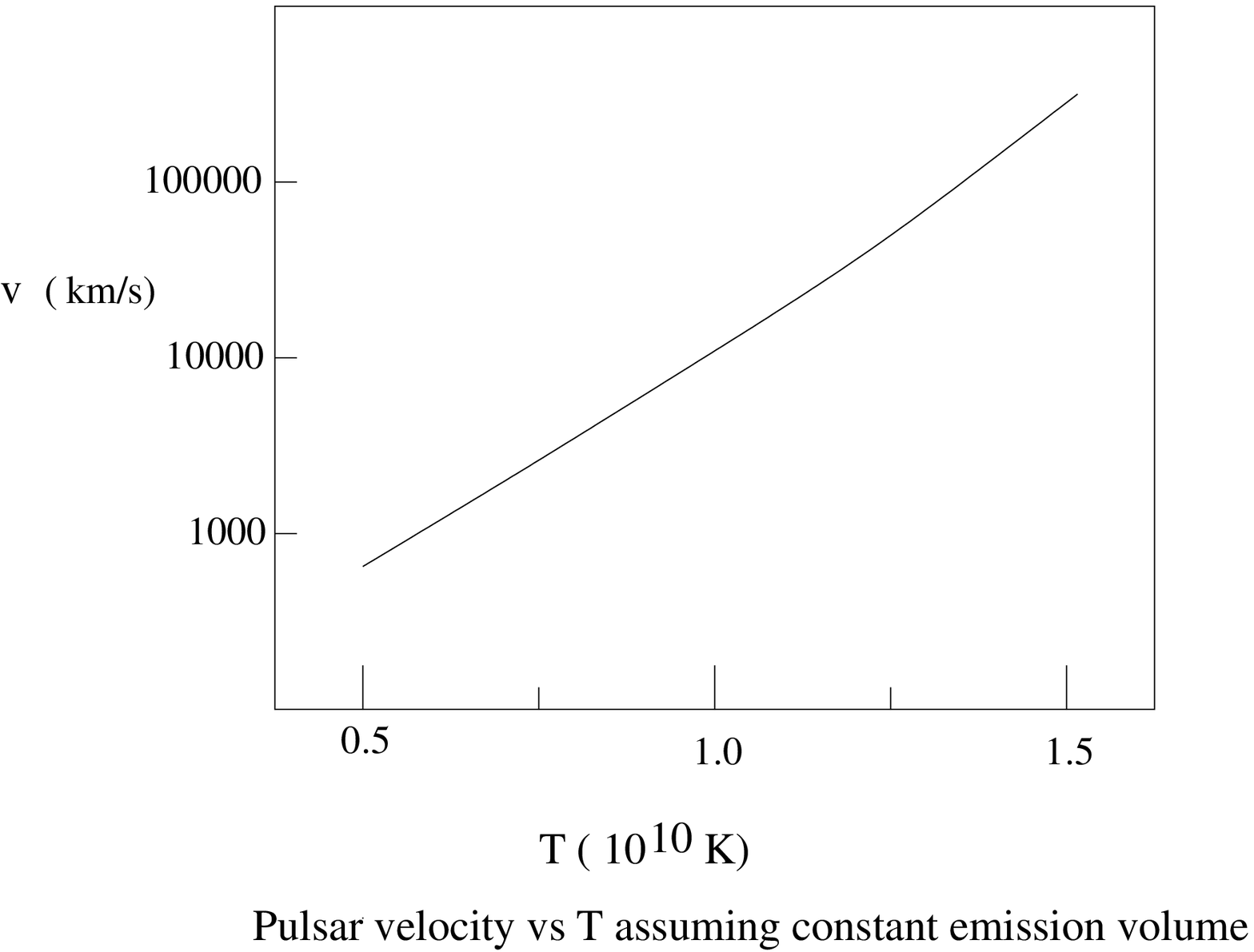,height=5cm,width=10cm}
%\caption{}
{\label{Fig.4}}
\end{center}
\end{figure}
 \clearpage
\vspace{-2cm}

\section{CONCLUSIONS}
   At the time the radius of the neutrinosphere is that of the neutron star
the modified URCA process dominates the energy emission.
\vspace{2mm}

   At this time the electrons produced in the modified URCA process are
in Landau levels, causing asymmetry of all neutrinos emitted.
\vspace{2mm}

  When the radius of the neutrinosphere is about 9.96 km vs the neutron
star radius of 10 km, the neutron star recives a kick of 1000 km and more,
consistent with observation
\vspace{2mm}

WE PREDICT A STRONG CORRELATION BETWEEN THE PULSAR'S VELOCITY AND
LUMINOSCITY FOR LARGE PULSAR KICKS, AS OBSERVED  (see Hansen et al)

\begin{figure}[h]
\begin{center}
\epsfig{file=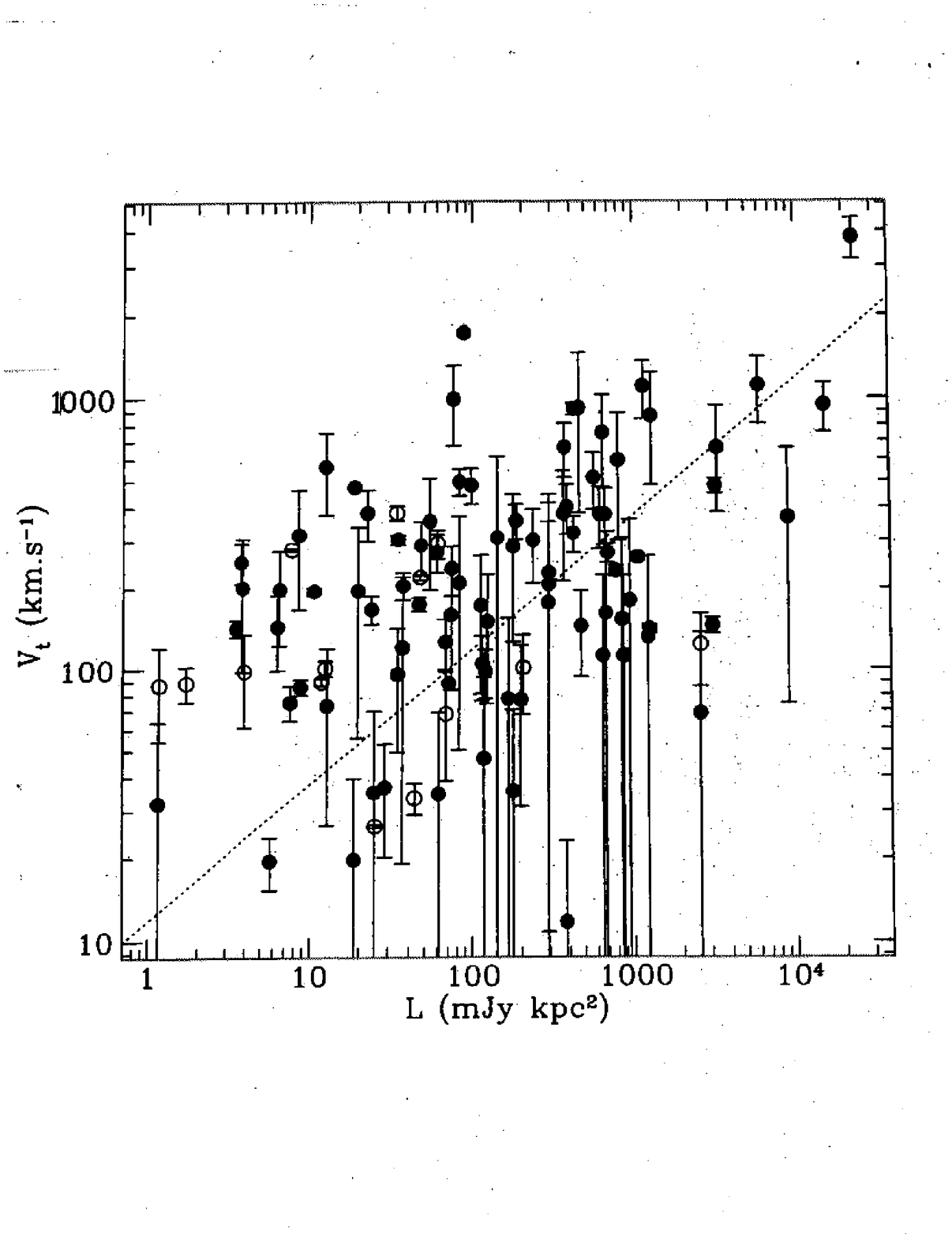,height=6cm,width=8cm}
%\caption{Pulsar Speed vs. Luminoscity}
%\label
\end{center}
\end{figure}
\vspace{-2cm}
WE PREDICT THE BIG KICK STARTS AT ABOUT 10 s, CONSISTENT WITH AN 
ANALYSIS  OF SN1987 (K. Hirata et al, P.R.D38, 448 (1988)).

\begin{figure}[h!]
\begin{center}
\epsfig{file=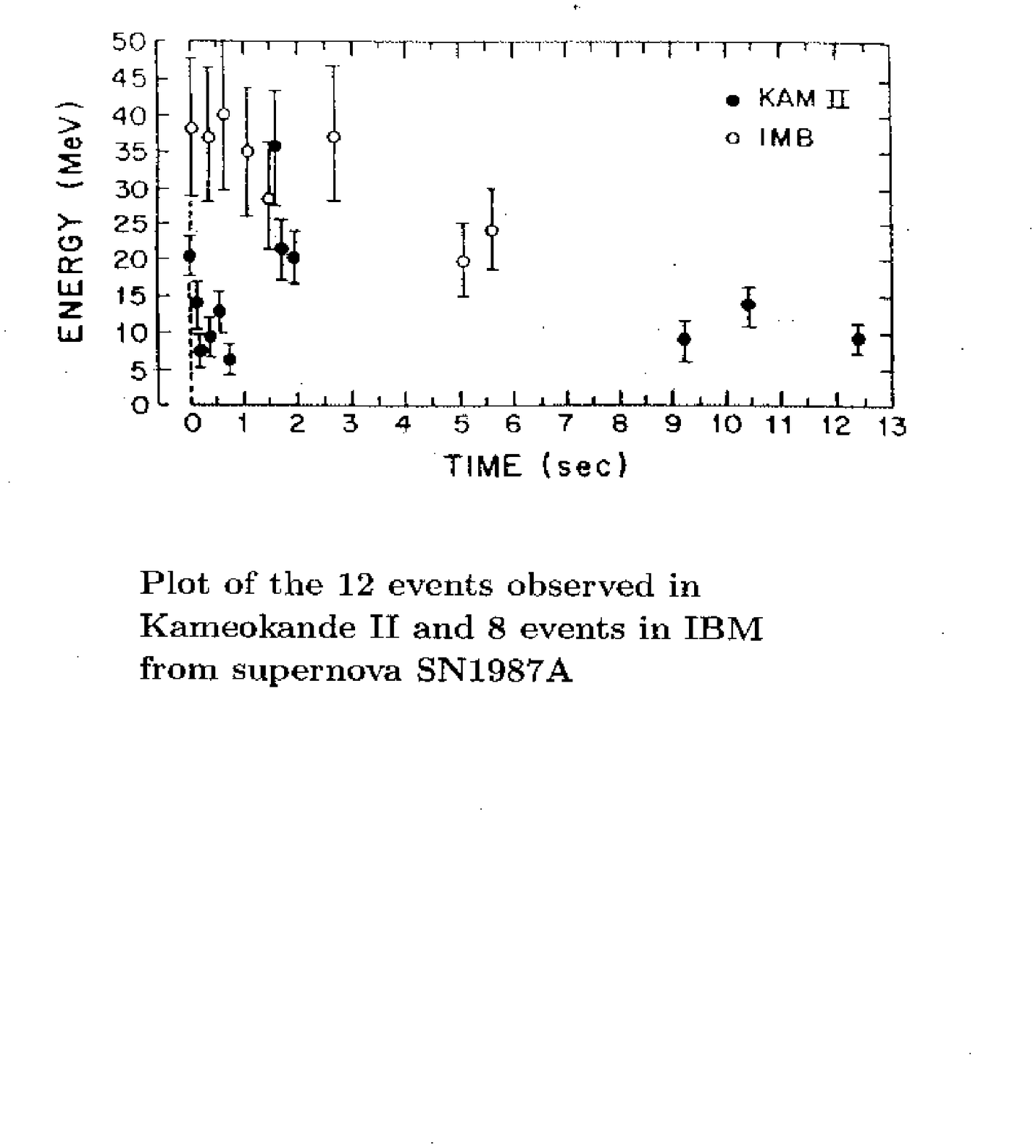,height=6cm,width=8cm}
%\caption{}
%\label
\end{center}
\end{figure}
\vspace{-2cm}

Acknowledgements: This work was supported in part by DOE contracts 
W-7405-ENG-36 and DE-FG02-97ER41014. We thank Sanjay Reddy and 
Richard Schirato for helpful discussions.

\end{document}